\documentclass[preprint,showpacs,showkeys]{revtex4}
\usepackage{epsf,psfig,pstricks,pst-grad}
\usepackage{amsmath}
\usepackage{amssymb}
\usepackage{epic}

\begin{document}
\def\Z{\mathbb{Z}}
\def\R{\mathbb{R}}
\def\k{{\mathbf k}}
\def\A{{\mathbf A}}

\title{The matrix rate of return}
\author{ Anna Zambrzycka}%
 \email{zanna12@wp.pl}
\affiliation{%
Institute of Physics, University of Silesia, Uniwersytecka
4, Pl 40007 Katowice, Poland.}%
\author{Edward W.~Piotrowski}\email{ep@alpha.uwb.edu.pl}\homepage{http://alpha.uwb.edu.pl/ep/sj/index.shtml}
 \affiliation{Institute of Mathematics,
University of Bia\l ystok, Lipowa 41, Pl 15424 Bia\l ystok,
Poland.}

\date{\today}

\begin{abstract}
 In this paper we give definitions of matrix rates of return which do not depend on the choice of basis
 describing baskets. We give their economic interpretation.
 The matrix rate of return describes baskets of arbitrary type and extends portfolio analysis to the complex variable domain.
 This allows us for simultaneous analysis of evolution of  baskets parameterized by complex variables in both continuous and discrete time models.
 \end{abstract}
\maketitle
\section{Introduction}
${}$\indent The goal of capital investment is the maximization of
a profit and minimization of possible losses. This goal cannot be
achieved by investing of the whole capital in the most profitable
enterprises.  Such situations do not happen. The future profit of
a market investment is uncertain, therefore, the investor creates
composite baskets consisting of capital investments of possibly
most diversified  character. This kind of procedure diversifies
the risk of enterprise. The description of the evolution of
multidimensional capital of this kind is essential to quantitative
analysis of the correlations related
with investment processes, in particular, these for which the
tools of traditional financial mathematics are unapplicable.
Depending on the point of view, in every capital basket we can,
besides of quantitative changes of individual components, observe
the flows between its components. The flows of capital can be
recorded even if no decisions about capital operations are taken.
Such situations require matrix description and enforce  the
generalizations of the calculus of interest rates in such a way
that they are sensitive to both quantitative changes of individual
elements and flows between them. This situation leads inevitably
to the matrix generalization of the interest rate calculus, see
\footnote[1]{E.~W.~Piotrowski,
 {\em The matrix rate of return} (in Polish), Przegl{\c a}d Statystyczny, {\bf 46/3} (1999), \;339-352.}.

\section{Linear homogeneous capital process}
Let us consider a capital of a banker which is a mixture of two
elements: $k_1$ -- an amount lent to client, $k_2$ -- the remaining
of assets. Let the variable $l\negthinspace\in\negthinspace \Z$ denote arbitrary time
interval. The process of change of the banker's capital related
with the amounts $k_1$ and $k_2$ can be described as follows
\renewcommand{\labelenumi}{\Alph{enumi}}
\begin{description}
\item[\textbf{$k_1 (l)$:}]According to the rate of interest  $\alpha_{1}(l)$ at which the banker
gave credit, the growth of the component $k_1$ is equal to
$\alpha_1 (l)k_1(l)$. In addition the amount of unreturned part of
credit decreases by the value  of repayment $\beta (l)k_1 (l)$
which is determined by the rate $\beta (l)$.
\item[\textbf{$k_2 (l)$:}] The capital $k_2$ increases by the amount of repayment of credit
$\beta (l)k_1 (l)$. Being in addition placed in, for example
liquid stocks of annual rate of return $\alpha_2 (l)$, it grows by
the amount $\alpha_2 (l)k_2(l)$.
\end{description}
The components of the banker's capital do form a basket
$k(l)=(k_1(l),k_2(l))$. It is represented by an element of the two
dimensional real vector space $\mathbb{R}^2$. The evolution of
this basket during the time $l\in \Z$ can be described by the
system of equations
\begin{equation}
\left\{
\begin{array}{ll}
     \hbox{$k_1 (l+1)=(1+\alpha_1 (l)-\beta(l))\;k_1 (l)$} \\
     \hbox{$k_2 (l+1)=\qquad \qquad \qquad \beta(l)\;k_1 (l) +(1+\alpha_2 (l))\;k_2 (l)$} \\
\end{array}
\right.
\label{1}.
\end{equation}
We interpret the negative values of components of the basket $k_m$,
$m\negthinspace=\negthinspace1,2,\ldots $  as debts of the banker.

The nonlinear (with respect to the remaining debt)  repayment
rates, that is for example the costs of service of credit or
taxes,  can  be presented in a form of linear repayment, after
appropriate modification of the factor $\beta (l)$. In particular
various borrower's obligations to the banker can be taken into
account with the help of the expression in variable proportion to
the amount of paid-off credit. This enables us to apply our
formalism in much wider context. This  evolution of the capital
can be described in such a way that any changes of components of
the basket are expressed as percent changes of those components.
In the vector space of baskets we can choose a new basis such that
we do not observe any flows of capital, but autonomous growth of
individual components only. Let us assume that $\alpha_1 (l)\neq
\alpha_2 (l)$ and $k_1^{'}(l)$  is as in previous basis, the
amount of the capital lent to client $k_1^{'}(l) =k_1 (l)$\,.
$k_2^{'}(l)$  is the sum of $\beta (l)$ part of a loan and
$\alpha_2 (l)-\alpha_1 (l)+ \beta (l)$ part of remaining banker's
assets. That is $k_2^{'}(l) =\beta (l)k_1(l)+(\alpha_2
(l)-\alpha_1 (l)+\beta(l))k_2(l)$. In these new variables
equations (\ref{1}), describing the changes in the basket separate
and take the following form: $$ \left\{
\begin{array}{ll}
     \hbox{$k_1^{'} (l+1)=(1+\alpha_1 (l)-\beta(l))\;k_1^{'} (l)$} \\
     \hbox{$k_2^{'} (l+1)=\qquad \:\:\:\: (1+\alpha_2 (l))\;k_2^{'} (l)$} \\
\end{array}\right. .
$$ In this way we do not observe any flows of capital and the
components $k_1^{'} $ and $k_2^{'} $ of the basket grow according
to the rates of interest $\alpha_1 (l)-\beta(l)$ and $\alpha_2
(l)$ for $k_1^{'} $ and $k_2^{'} $ respectively. The formalism of
financial mathematics should not depend on choice of basis to
describe the baskets.  It seems that the matrix rates presented
below are necessary to obtain the basis independent description of
the economical reality.

\section{The matrix rate of return}
Capital processes are described by linear homogeneous difference
equations. In the matrix notation they take the form:
\begin{equation}
\k (l+1)-\k (l) = \underline{\mathbf{R}}(l)\, \k (l)\,, \;\;\; \text{hence}\;\;\; \k (l+1) = (\mathbf{I} + \underline{\mathbf{R}}(l))\, \k (l)\,,\label{3}
\end{equation}
where $\k (l)\negthinspace\in\negthinspace \R^{M}$, $\underline{\mathbf{R}}(l)$ and
$\mathbf{I}$ are real matrices of dimension $M\negthinspace\times\negthinspace M$.
$\mathbf{I}$ denotes the unit matrix. Matrix
$\underline{\mathbf{R}}(l)$ is called the matrix credit rate of
return or the matrix lower rate. The adjective lower refers to the
way of the description of the growth of the vector $\k (l)$. These
changes are expressed as the effects of linear transformations
applied to the vector $\k (l)$ of the basket  in the moment $l$
prior to this change. The space $\R^{M}$ of all possible baskets,
is called the phase space of baskets and linear homogeneous
difference or differential systems of equations of first order are
called equations of motion of the basket. For
$M\negthinspace=\negthinspace2$  the matrix
$\underline{\mathbf{R}}(l)$  generating the evolution $(\ref{1})$
equals to
\begin{equation}\label{m}
\underline{\mathbf{R}}(l)=\left( \begin{array}{cc}
\alpha_{1}(l)-\beta(l) & 0 \\
\beta(l) & \alpha_{2}(l)
\end{array}
\right).
\end{equation}
If the state of the basket in the initial moment $p$ equals to $\k
(p)$, then the solution of the evolution  equation (\ref{3}) takes
the form
\begin{equation}
\k(r)=\left( \mathcal{T} \prod_{s=p}^{r-1}\left( \mathbf{I}+\underline{\mathbf{R}}(s)\right) \right)\k(p)\,,\label{k}
\end{equation}
where $\mathcal{T}$ denotes the chronological ordering operator.
It orders all matrices chronologically: the matrices  with later
arguments are moved to the left of the matrices with earlier
arguments, that is $$ \mathcal{T} \prod_{s=p}^{p+n} \A
(s)=\A(p+n)\A (p+(n-1))\ldots \A (p+1)\A (p)\,, $$ where $\A (s)$ is
an arbitrary sequence of matrices.

The chronological product $\mathcal{T} \prod_{s=p}^{r-1}\left(
\mathbf{I}+\underline{\mathbf{R}}(s)\right)$ contains more
detailed information about changes of capital in the basket than
the usually used  quotients $\frac{\sum_{m=1}^{M}k_m
(r)}{\sum_{m=1}^{M}k_m (p)}$\,. The possibility to use the standard
calculus of interest rates depends on way of description of the
basket, that is on choice of a basis in our language. For this
reason, in order to obtain an invariant description,  the rate of
growth $\underline{\mathbf{R}}(l)$ is promoted to a composite
object with definite  transformation rules corresponding to the
changes of reference frame (an observer).

Let us consider processes for which the matrix $\mathbf{I} +
\underline{\mathbf{R}}(l)$ is nonsingular. Then, introducing the
concept of the matrix discount  rate of return or the matrix upper
rate $\overline{\mathbf{R}}(l)$, it is possible rewrite the
equation of motion (\ref{3}) in the following form:
\begin{equation}
\k (l+1)-\k(l)=\overline{\mathbf{R}}(l)\,\k (l+1)\,. \label{3g}
\end{equation}
Comparing formulae (\ref{3}) and (\ref{3g}) we obtain the relation
between both types of  matrix rates introduced above:
\begin{equation}
\left( \mathbf{I}+\underline{\mathbf{R}}(l)\right)\left( \mathbf{I}-\overline{\mathbf{R}}(l)\right)= \left( \mathbf{I}-\overline{\mathbf{R}}(l)\right)\left( \mathbf{I}+\underline{\mathbf{R}}(l)\right)=\mathbf{I}\, .\label{zw}
\end{equation}
Solving the above equation with respect to
$\underline{\mathbf{R}}(l) $ we obtain
\begin{equation}
\underline{\mathbf{R}}(l)=\overline{\mathbf{R}}(l)+\overline{\mathbf{R}}^2 (l)+\overline{\mathbf{R}}^3 (l)+\ldots .\label{2.7}
\end{equation}
From the point of view  of capitalisation from upper, we can
interpret the above formulae as the contribution of discount rate
into increase of capital by summation of all interests from
interests (the geometrical sequence). According to Eq. (\ref{2.7})
the formulae for $\overline{\mathbf{R}}(l)$ takes the form $
-\overline{\mathbf{R}}(l)=(-\underline{\mathbf{R}}(l))+(-\underline{\mathbf{R}}(l))^2
+(-\underline{\mathbf{R}}(l))^3 +\ldots\, . $
    Notice that for fixed
argument $l$ the matrix rates are commuting.

When we perform the formal change of the direction of time the
credit and discount rate change their signs and they change their
roles. Therefore, the formulae containing these matrices are
symmetric with respect to the time reflection. The appropriate
formulae for (\ref{k}) can be obtained with the help of (\ref{zw})
as the solution of the equation of motion:
\begin{equation}
\k(p)=\left( \mathcal{T}^{'} \prod_{s=p}^{r-1}\left( \mathbf{I}-\overline{\mathbf{R}}(s)\right) \right)\k(r)\,, \label{2.8}
\end{equation}
where $\mathcal{T}^{'}$ is the antichronological operator which
orders matrix rates in the direction opposite  to that
corresponding to the operator $\mathcal{T}$.

If the matrix rates at different moments of time commute, that is
in the case of $M\negthinspace =\negthinspace 1$ or for any $M$ if
matrix rates are time-independent, we can neglect the operators
$\mathcal{T}$ and $\mathcal{T}^{'}$ in the solutions (\ref{k}) and
(\ref{2.8}).

\section{Interpretation the matrix rate of return}
The matrix rate of return $\underline{\mathbf{R}}(l)=\left(
\underline{\mathbf{R}}_{mn}(l)\right)$ can be given as sum of two
matrices
    $
\underline{\mathbf{R}}(l)=\underline{\mathbf{C}}(l)+\underline{\mathbf{D}}(l),
    $
with $\underline{\mathbf{C}} (l)$  the matrix of flows (the name
is justified by the property that sum of the elements of each
column is equal zero) and $\underline{\mathbf{D}}(l)$ -- the
diagonal matrix called the matrix of growths. This decomposition
is unique, when the basis is fixed. Introduction of the matrix
rate of return is essential, when we cannot transform the matrix
$\underline{\mathbf{C}}(l)$ to matrix zero. Moreover, the
off-diagonal elements of the matrix $\underline{\mathbf{R}}(l) $,
$\underline{\mathbf{R}}_{mn}(l)$ for
$m\negthinspace\neq\negthinspace n$, see Eq. $(\ref{m})$, describe
which part of capital of the $n$-th component of the basket flows
to the $m$-th component of the basket. Diagonal elements
$\mathbf{R}_{mm}(l)$ describe the growths. 
The rate of growth of the $m$-th component of
the basket equals $\alpha_{m}(l)=\mathbf{R}_{mm}(l)-\sum_{m\neq
n}\underline{\mathbf{R}}_{nm}(l)$, that is, the diagonal element
corrected by all out-flows of the capitals related to the
component $\k_{m}(l)$.

\subsection{Example}
\begin{description}
\item[(i)]Let us restrict the process from section II to  the
cases when $\beta (l)=0$ and the coefficients $\alpha_i$ do not
depend on time $l$ and  the second coefficient  is two times
bigger than the first  one. Then the matrix lower  rate is equal
to $ \underline{\mathbf{R}}=\left( \begin{array}{cc}
\alpha & 0 \\
0 & 2\alpha
\end{array}
\right)\negthinspace.
$
According  to the principal formula (\ref{zw}) the corresponding
matrix upper  rate is given by:
    $ \overline{\mathbf{R}}=\left(
\begin{array}{cc} \frac{\alpha}{1+\alpha} & 0 \\ 0 &
\frac{2\alpha}{1+2\alpha}
\end{array}
\right)\negthinspace.
    $
The process contains the matrix of growths only because the matrix
of flows is zero. It is possible to analyze the profits on the
grounds of autonomic evolution in one-dimensional subspaces of the
phase space --  the classical concept of interest rates is
applicable here. \item[(ii)]We can look at this process in a
different way. Namely by describing it in the coordinates in the
new basis in the space of baskets. Let the reference basis consist
of  client's debts and all banker's capital assets. Then the
equation of motion of the basket (\ref{1}) takes the form
\begin{equation}
\left\{\begin{array}{ll}
     \hbox{$\widetilde{\k}_{1} (l+1)=(1+\alpha)\;\widetilde{\k}_{1} (l)$} \\
     \hbox{$\widetilde{\k}_{2} (l+1)=\quad \;\;-\alpha\;\widetilde{\k}_{1}(l)+(1+2\alpha )\;\widetilde{\k}_{2}(l)$}. \\
\end{array}\right.\label{3.2}
\end{equation}
The matrix of flows determined by the above equation is non zero
now. The matrix of flows and the matrix of growths equals to
    $$
\underline{\mathbf{C}}=\left( \begin{array}{cc} \alpha & 0 \\
-\alpha & 0
\end{array}
\right), \qquad \underline{\mathbf{D}}=\left( \begin{array}{cc}
0 & 0 \\
0 & 2\alpha
\end{array}
\right),
    $$
respectively.
    The debt of a client $\tilde{k}_{1}$ changes in a similar way to the previous example,
    though not as result of an autonomic growth, but due to an outflow of the banker's capital.
    Banker's  capital grows according to the same rate as investment $k_2 (l)$.
\item[(iii)] In the  basis  given by  Eq. (\ref{zw}) the
 matrix upper rate for the equation of motion (\ref{3.2}) takes the
form
    $$ \overline{\mathbf{R}}=\left( \begin{array}{cc}
\frac{\alpha}{1+\alpha} & 0 \\
-\frac{\alpha}{(1+\alpha)(1+2\alpha)} & \frac{2\alpha}{1+2\alpha}
\end{array}
\right) $$ and it is the sum of the following matrices of flows
and growths
    $$ \overline{\mathbf{C}}=\left(
\begin{array}{cc} \frac{\alpha}{(1+\alpha)(1+2\alpha)} & 0 \\
-\frac{\alpha}{(1+\alpha)(1+2\alpha)} & 0
\end{array}
\right), \qquad \overline{\mathbf{D}}=\left( \begin{array}{cc}
\frac{2\alpha ^2}{(1+\alpha)(1+2\alpha)} & 0 \\
0 & \frac{2\alpha}{1+2\alpha}
\end{array}
\right).
$$
\end{description}
Different points of view at the same capital process presented in
(i), (ii), (ii) are equally correct and sensible. Comparing the
credit convention in variant (ii) with the discount convention
(iii) we can note the essential difference between the matrices
$\underline{\mathbf{D}}$ and $\overline{\mathbf{D}}$. Growth of
first component is only the effect of flows in the first case
while in second case this component has partial autonomy in its
growth. The indicated difference in interpretation can be a reason
of many financial embezzlement, exactly in the same way as it
happens in case of  nonpayment of interests of overdue interests
in simple capitalisation. We will refer the asymmetry of this kind
in the description of flows as the paradox of difference rates.

The interpretation of the matrix rate $\overline{\mathbf{R}}$ is
analogous to that of the matrix $\underline{\mathbf{R}}$ with the
only difference that the rate $\overline{\mathbf{R}}$ defines the
capital changes with respect to the oncoming moment of the
settlement of accounts. The convention of discount replaces the
convention of credit. The free choice of convention is not
possible for nonreversible processes i.e\mbox{.} when the matrices
$\underline{\mathbf{R}}$ and $\overline{\mathbf{R}}$ are singular.

\section{The formalism of continuous description of credit}
For the capital calculus to be transparent  and readable for
practitioners the formal solution of the equation of motion
(\ref{3}) can be modeled numerically or presented it in dense form
with the help of the limiting process which transforms  the
discrete models described by linear difference equations into
continuous ones with differential equations. Assume that we
consider time scales such that the periods of time between the
changes of components of the basket is infinitesimal and  equals
to $\tau = t_{l+1}-t_l$. After rescaling of the time domain of the
basket the equation of motion (\ref{3}) takes the form:
    $$
\frac{\k (t_l +\tau)-\k(t_l
)}{\tau}=\frac{\underline{\mathbf{R}}(l)}{\tau}\,\k (t_l )\,.
    $$
In the limit $\tau \rightarrow 0$ we obtain
\begin{equation}
\frac{d \k (t)}{dt}=\mathbf{R} (t)\, \k(t)\;\;\;\;\text{where}\;\;\;\;\mathbf{R} (t):=\lim_{\tau \rightarrow 0}\frac{\underline{\mathbf{R}}(l)}{\tau }\vert_{t=t_l}.\label{4}
\end{equation}
Matrix $\mathbf{R} (t)$ is called the differential matrix rate of return.


The formal solution of the equation (\ref{4}) is given by the formula
\begin{equation}
\k(t) = \left( \mathcal{T} \text{e}^{\int_{t_{0}}^{t}\mathbf{R} (t^{'}) dt^{'}}\right)\k(t_{0} )\,.\label{4.2}
\end{equation}
The chronologically ordered exponential function is infinite
series in the differential matrix rate
    $$ \mathcal{T}
\text{e}^{\int_{t_{0}}^{t}\mathbf{R} (t^{'})
dt^{'}}=\mathbf{I}+\int_{t_{0}}^{t}\mathbf{R} (t_1 ) dt_1
+\int_{t_{0}}^{t}\mathbf{R} (t_1 )\int_{t_{0}}^{t_1 }\mathbf{R}
(t_2 )dt_2 dt_1 +\ldots .
    $$
If the differential matrix rate is constant, the chronological
operator $\mathcal{T}$ is the identity. The expression on the
right hand side of the equation (\ref{4.2}) which describes the
time evolution of the basket becomes transformed to the standard
matrix exponential function
    $ \mathcal{T}
\text{e}^{\int_{t_{0}}^{t}\mathbf{R} (t^{'}) dt^{'}}\negthinspace=\text{e}^{(t-t_0
)\mathbf{R}_{0}}.
    $

\section{Complex rate of return}
Let us consider baskets which have oscillating components. Part of
their capital becomes a debt at some moment and a desirable good
at another time. This phenomenon is called pumping of capital in
the language of financial market.  Evolution of basket of this
kind is described by one of the most popular model of physics --
the harmonic oscillator. Let us consider the differential matrix
rate $$ \mathbf{R}(t)=\left(
\begin{array}{cc}
 0 & -b \\
b & \;0
\end{array}\right)=\left( \begin{array}{cc}
 b & \;0 \\
0 & -b
\end{array}
\right)+\left( \begin{array}{cc}
 -b & -b \\
b & b
\end{array}
\right). $$ It is easy to imagine the corresponding bank contracts
leading to the flows of capital in basket and  autonomic increases
of components of the basket generated by these flows. Let us
consider the complex extension $\mathbb{C}^{2}$ of the phase space
$\mathbb{R}^{2}$. Then $\widetilde{\k}_{1}=(1,\text{i})$ and
$\widetilde{\k}_{2}=(1,-\text{i})=\widetilde{\k}_{1}^{*}$ are the
 eigenvectors of the matrix rate $\mathbf{R}(t)$ with eigenvalues $-\text{i}\,b$ and $\text{i}\,b$ respectively.
The description of process is simplified because the basket splits
into two independent components with abstract complex capital.
Although that baskets of the complex capital are the abstract
concepts, they have the interpretation in the real basis due to
matrix of transition. The equation of motion of the basket in the
basis of eigenvectors  $\{\widetilde{\k}_{1},\widetilde{\k}_{2}\}$
takes the form: $$
\widetilde{\k}_{1}(t)=\text{e}^{-\text{i}b(t-t_{0})}\,\widetilde{\k}_{1}(t_0
)\;\;\;,\;\;\;\widetilde{\k}_{2}(t)=\text{e}^{\text{i}b(t-t_{0})}\,\widetilde{\k}_{2}(t_0
)\,. $$ 
In the initial, real basis one has
    $$ \k (t)=\left(
\begin{array}{cc}
 \cos (b(t-t_{0})) & -\sin (b(t-t_{0})) \\
\sin (b(t-t_{0})) & \phantom{-}\cos (b(t-t_{0}))
\end{array} \right) \k (t_0)\,.
    $$
This equation describes motion along the circle centered  at the
beginning of the cartesian coordinates of the basket. The period
of return to the same point of the phase space equals to
$T=\frac{2\pi }{\vert b\vert}$.

\section{The indefinite matrix rate}
We define the indefinite logarithm, see \footnote[2]{M.~P.~Frank,
{\em The Indefinite Logarithm, Logarithmic Units, and the Nature
of entropy}\/, arXiv:physics/0506128 v 1, (2005).}, to be a
mathematical object representing the abstract concept of the
logarithm with an unfixed base. For any given real number $x\negthinspace>\negthinspace0$,
the indefinite logarithm of $x$ written as $[\log x]$, is a
special type of mathematical object called a logarithmic quantity
object, which we define as follows [2]:
$$ [\log x]:=\left\lbrace (b,y)\vert b > 0, y=\log_{\,b}
x\right\rbrace. $$ Indefinite logarithmic quantities are
inherently scale-free objects, that is, they are non-scalar
quantities and they can serve as a basis for logarithmic spaces,
which are natural systems of logarithmic units suitable for
measuring any quantity defined on a logarithmic scale. Although
the above definition is restricted to positive real numbers, it
could be extended to non-zero complex numbers too. The concept of
the rate of interest is connected with the time scale. To get rid
of explicit time scale dependence we introduce the indefinite
matrix rate, which is the generalization of the indefinite
logarithm to the multidimensional case: $$ \left[ \log \left(
\mathcal{T} \text{e}^{\int_{t_{0}}^{t}\mathbf{R} (t^{'})
dt^{'}}\right) \right]. $$ In the above definition the argument of
the logarithm is a matrix.

\section{Conclusions}

Every matrix can be deformed to a diagonalizable complex matrix by
arbitrarily small deformations of their  elements \footnote[3]{G.
Arfken, {\em Diagonalization of Matrices}, Mathematical Methods
for Physicists, Academic Press (1985), 217-229.}\footnote[4]{A.~P.
Mishina, I.~V. Proskuryakov, {\em Higher algebra. Linear algebra,
polynomials, general algebra}, Pergamon  (1965)}, that is the set
of diagonalizable linear transformations of the
 complexified phase space $\mathbb{C}^{M}$ is dense  in the space all linear maps of $\mathbb{C}^{M}$.
From the above it follows that, the evolution of every capital
basket can be represented as set of non interacting complex
capital investments. Therefore, in the complex extended phase
space the  decomposition of the matrix rate
$\underline{\mathbf{R}}(t)=\underline{\mathbf{C}}(t)+\underline{\mathbf{D}}(t)$
can always be done in such a way that the matrix of flows is zero.
As in traditional financial mathematics the real parts of nonzero
elements of a diagonal matrix rate measure the loss or gain of
complex investments. The imaginary parts inform about periodicity
of changes in proportion between real component of complex
investments and its imaginary partner. To every  periodic changes
of proportion of the elements of the complex components of capital
there will correspond its complex conjugated partner. The
evolution of real capital of basket is  most easily observed in
terms of its components with respect to the basis of eigenvectors
of matrix rates.

    By the  choice of appropriate moments of entering
and exit from capital process  the oscillations like above can be
used as particularly effective mechanism of enlarging  of the
capital giving the similar results as financial leverage.
\begin{acknowledgements} 
We are greatly indebted to prof.~Zbigniew Hasiewicz for helpful remarks.
\end{acknowledgements}



\end{document}